\documentclass[aip,jap,a4,reprint,twoside,floatfix,superscriptaddress]{revtex4-1} 

\pdfoutput=1
\usepackage{graphicx}
\usepackage{amsmath}
\usepackage{amssymb}
\usepackage{amsfonts}
\usepackage{dcolumn}
\usepackage{epsfig}
\usepackage{subfigure}
\usepackage{bm}

\begin{document}

\title{\bf Determination of intrinsic ferroelectric polarization in lossy improper ferroelectric systems }

\author {Ujjal Chowdhury} \email{ujjalchowdhury1985@gmail.com} \affiliation {Nanostructured Materials Division, CSIR-Central Glass and Ceramic Research Institute, Kolkata 700032, India}
\author {Sudipta Goswami} \affiliation {Department of Solid State Physics, Indian Association for the Cultivation of Science, Kolkata 700032, India}
\author {Dipten Bhattacharya} \email{dipten@cgcri.res.in} \affiliation {Nanostructured Materials Division, CSIR-Central Glass and Ceramic Research Institute, Kolkata 700032, India}
\author {Arindam Midya} \affiliation {Saha Institute of Nuclear Physics, 1/AF Bidhannagar, Kolkata 700064, India}
\author {P. Mandal} \affiliation {Saha Institute of Nuclear Physics, 1/AF Bidhannagar, Kolkata 700064, India}

\date{\today}

\begin{abstract}
We measured the intrinsic hysteretic polarization in lossy improper and nanoferroelectric systems where the nonhysteretic polarization and leakage are large and the relaxation takes place over a broader time scale. We used different measurement protocols such as standard single triangular voltage pulse, a pulse train of PUND (Positive Up Negative Down), and an even more complicated pulse train of fourteen voltage pulses and compared the results obtained. We show that a protocol which sends a train of fourteen pulses is more appropriate for extracting relaxed (i.e., time scale independent) and intrinsic remanent polarization for these samples. We also point out that it is possible to select and design an appropriate measurement protocol depending on the magnitude of polarization and leakage of the system.  
\end{abstract}

\pacs{75.80.+q, 75.75.+a, 77.80.-e}
\maketitle

During the last decade and a half, the search for new multiferroic material with strong magnetoelectric coupling at room temperature remained at the forefront of research in magnetoelectric multiferroics.\cite{Dong} The possibility of observing strong magnetoelectric coupling is higher in "improper" ferroelectrics than in systems with "proper" ferroelectricity.\cite{Mostovoy} However, intrinsic ferroelectric polarization is nearly two to three orders of magnitude smaller in improper ferroelectrics. This weak 
polarization is often masked with large leakage and contributions from nonferroelectric polarization. Moreover, coxisting ferroelectric, nonferroelectric, and conducting regions as well as large depolarizing field at the sample-electrode interface give rise to complicated domain switching kinetics and broader relaxation time scale.\cite{Kim, Jo, Noh} Therefore, determination of relaxed and intrinsic ferroelectric polarization in them is extremely difficult.\cite{Scott, Feng, Catalan, Fina}
Depending on the resistivity and polarization of the sample, an appropriate technique needs to be developed in order to extract the tiny intrinsic and relaxed hysteretic or switchable polarization by eliminating stronger contributions from nonhysteretic (i.e., nonswitchable) and leakage components. The switchable polarization also consists of two components, namely, remanent and non-remanent. In this Letter, we report that we have measured the intrinsic remanent polarization by taking care of the relaxation charactertics and seperating out different contributions to the total polarization using an appropriate measurement protocol. The intrinsic remanent polarization turns out to be, expectedly, independent of measurement time scale. 

We have used three distinctly different profiles of voltage pulses in the Sawyer-Tower circuit. The first one is the widely used single triangular wave. The second one is PUND (Positive Up Negative Down) and the third one is a special type of pulse profile known as remanent polarization measurement protocol.\cite{Evans} For single triangular voltage pulse and remanent hysteresis measurement protocol, the loop tracer of Radiant Inc. was used while for PUND, the TF analyzer Aixaact was used. In addition, we have measured the polarization by pyroelectric current technique as well, using the Keithley sourcemeter 2400. The experiments were carried out on lossy improper and nanosized ferroelectric systems such as nanoscale BiFeO$_3$, orthorhombic LuFeO$_3$, and Pr$_{0.55}$Ca$_{0.45}$MnO$_3$. Silver paste and wires were used for electrical connections in two-probe configuration. We compared the results obtained from different techniques in order to find out the most suitable one for determining the relaxed (i.e., time-scale independent) and intrinsic ferroelectric polarization of the sample depending on the magnitude of its leakage and polarization.

\begin{figure}[ht!]
\centering
\includegraphics[width=110mm]{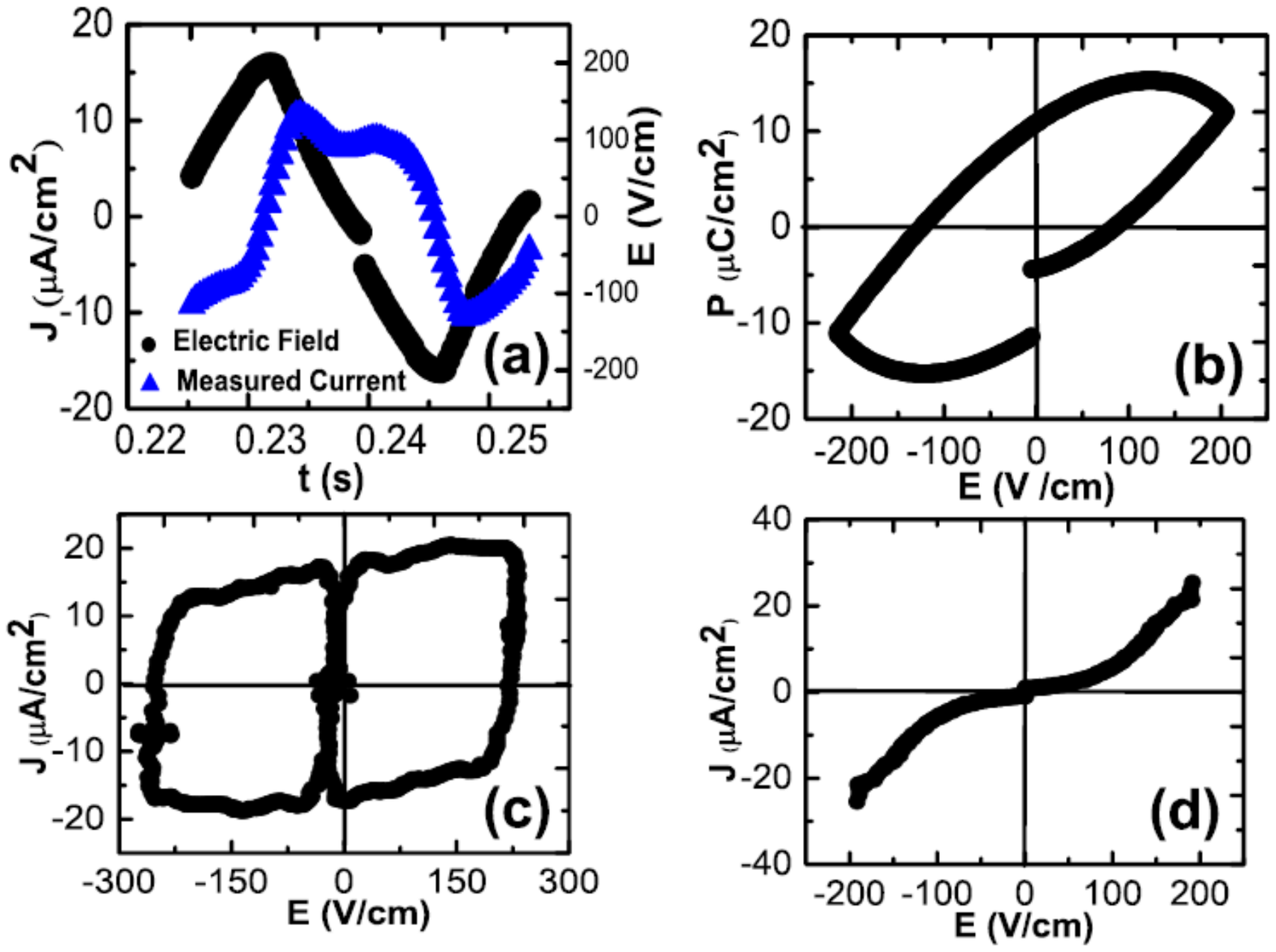}
\caption{(color online) The (a) pulse profile, current response, (b) P-E hysteresis loop, (c) polarization current versus field pattern, and (d) leakage current versus field pattern for orthorhombic LuFeO$_3$ corresponding to the single triangular pulse commonly used in Sawyer-Tower circuit for measuring ferroelectric polarization in proper ferroelectric systems. }
\label{overflow}
\end{figure}

\begin{figure}[hb!]
\centering
\includegraphics[width=110mm]{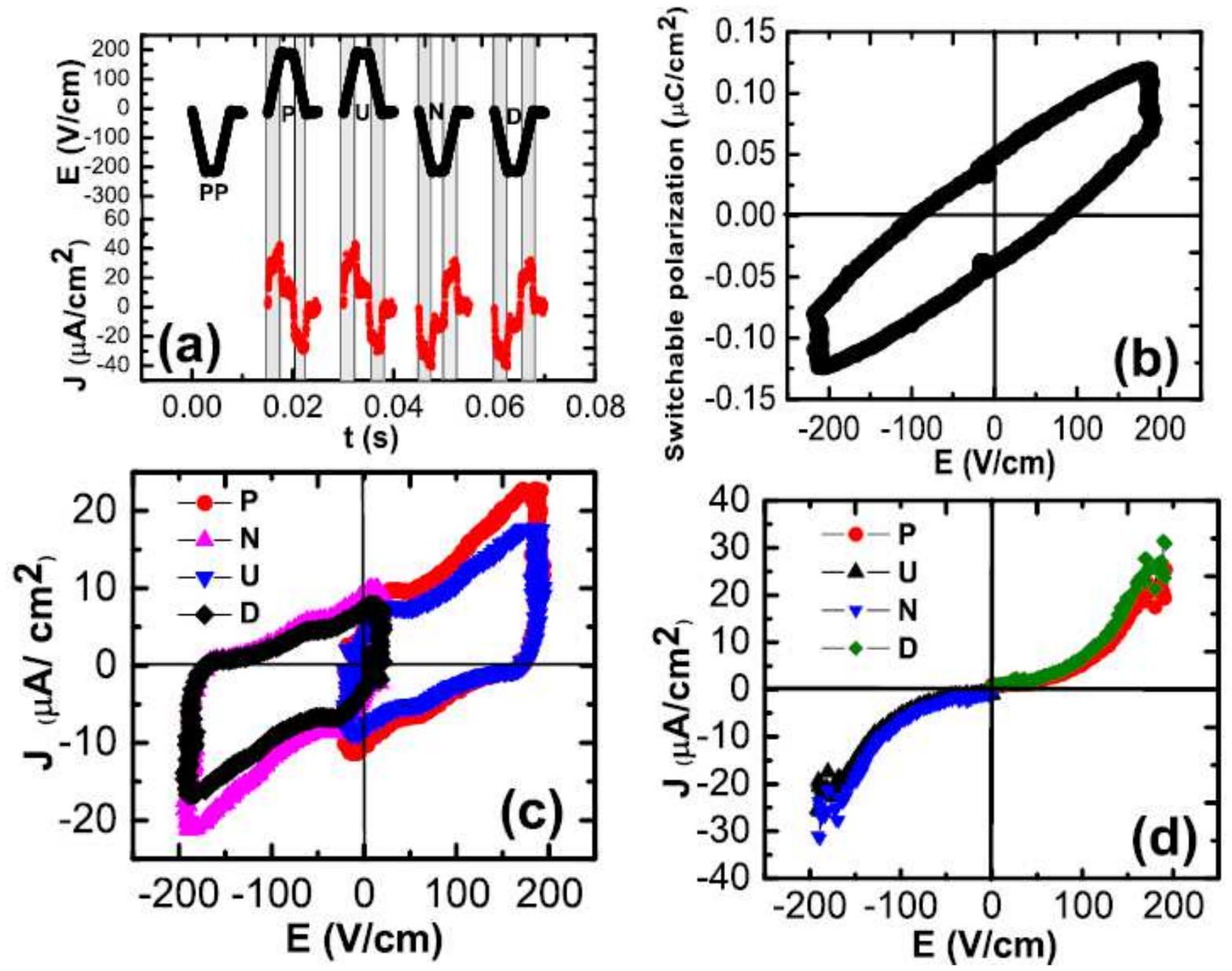}
\caption{(color online) The (a) pulse profile, current response, (b) P-E hysteresis loop, (c) polarization current versus field pattern, and (d) leakage current versus field pattern for orthorhombic LuFeO$_3$ corresponding to the PUND profile used for extracting intrinsic ferroelectric polarization in samples exhibiting small polarization. }
\label{overflow}
\end{figure}

\begin{figure}[ht!]
\centering
\includegraphics[width=65mm]{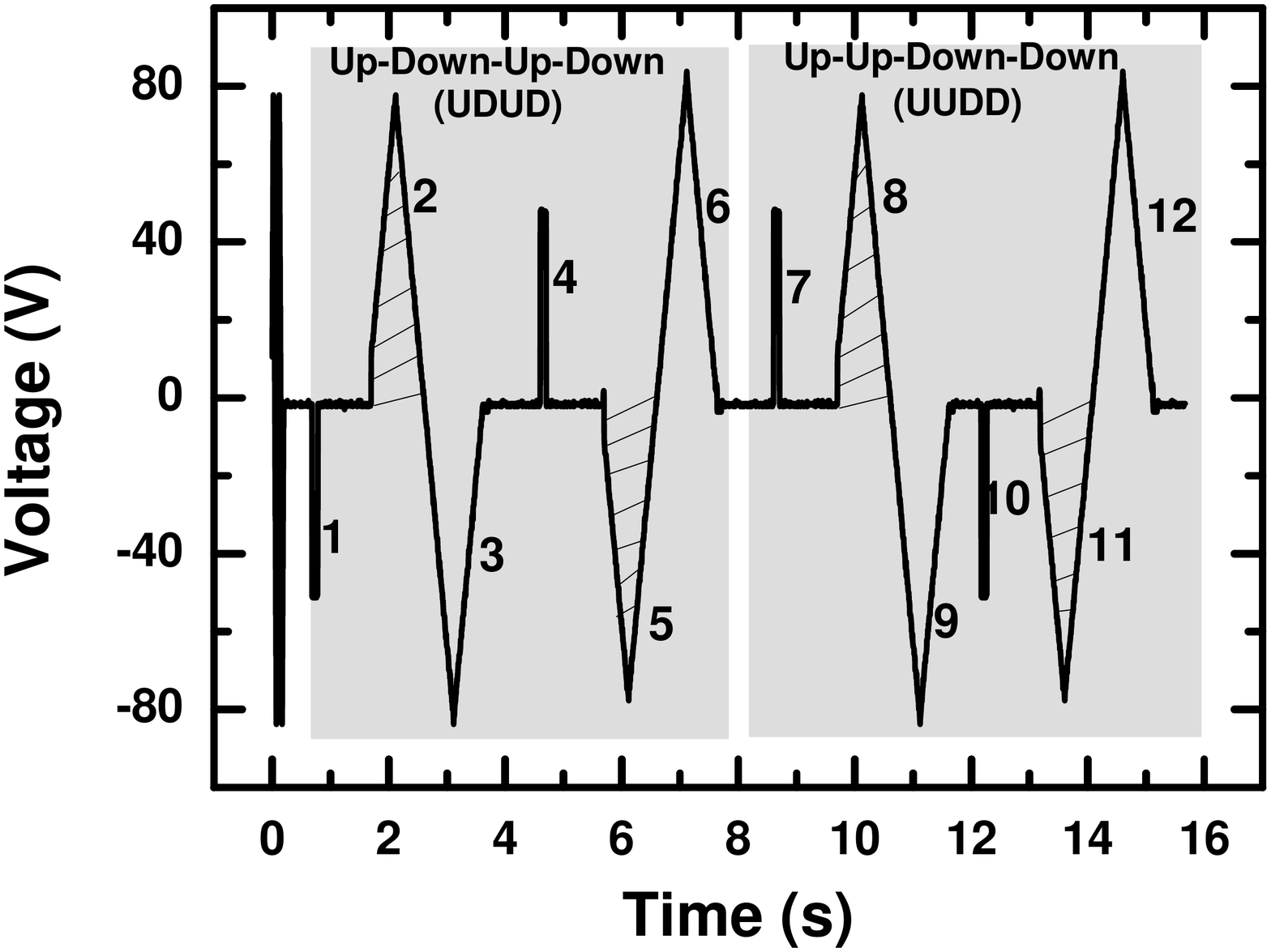}
\caption{(color online) The voltage pulse profile corresponding to the remanent hysteresis loop measurement protocol. After initial depolarization altogether 12 voltage pulses are sent to measure two complete loops (loop1 from UDUD pulses and loop0 from UUDD pulses). Subtraction of loop0 from loop1 gives the remanent hysteresis loop. }
\label{overflow}
\end{figure}

\begin{figure}[hb!]
\centering
\includegraphics[width=75mm]{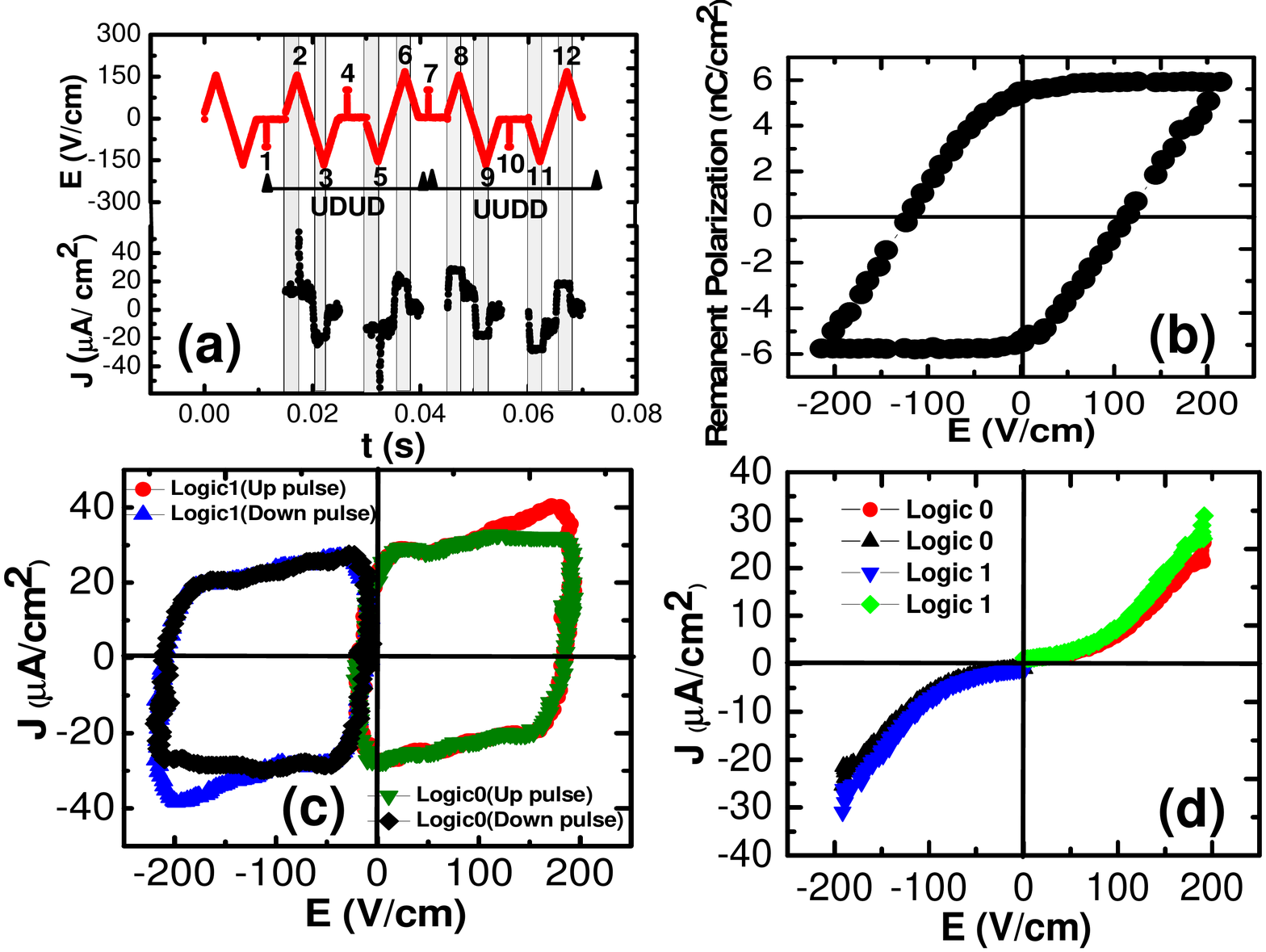}
\caption{(color online) The (a) pulse profile, current response, (b) remanent hysteresis loop, (c) polarization current versus field pattern, and (d) leakage current versus field pattern for orthorhombic LuFeO$_3$ corresponding to the specialized protocol employed for extracting the intrinsic remanent hysteresis loop. }
\label{overflow}
\end{figure}

In Fig. 1, we show the pulse profile, current response, and polarization ($P$) versus electric field ($E$) hysteresis loop for LuFeO$_3$ obtained from sending a single triangular pulse in the Sawyer-Tower circuit. This pulse simultaneously polarizes and measures the polarization and is most commonly used for measuring the ferroelectric hysteresis loop for proper ferroelectric systems exhibiting large polarization ($>$10 $\mu$C/cm$^2$). Quite clearly, it does not measure the contribution of ferroelectric and nonferroelectric polarization separately and, as a result, a cigar-shaped loop characteristic of lossy dielectric system is obtained for LuFeO$_3$.\cite{Scott} It is not possible to determine accurately the ferroelectric properties such as saturation and remanent polarization, coercivity etc. from this loop. Next a somewhat better and more useful protocol called PUND was used for extracting the intrinsic hysteretic polarization for lossy improper ferroelectrics.\cite{Rabe} In the case of PUND, a pulse train comprising of five separate voltage pulses is sent (Fig. 2) to measure the hysteretic and nonhysteretic polarizations. Eventually, contribution of nonhysteretic polarization is subtracted from the overall polarization to obtain the hysteretic polarization alone. In spite of its large popularity in recent time, we show that yet another and more involved protocol is needed to measure polarization as small as $\sim$1-10 nC/cm$^2$. In Fig. 3, we show the voltage pulse profile used for measuring remanent hysteresis loop. This protocol sends a train of fourteen voltage pulses and induces repeated switching and/or reinforcment in consecutive pulses. After two depolarizing pulses, it sends a polarizing pulse (pulse 1 in Fig. 3) which polarizes the domains along a certain direction. The next pulse (pulse 2) switches the domains by 180$^o$ and measures one half of the polarization loop. Next two pulses (3 and 4) switch the domains consecutively in opposite directions without carrying out any measurement. Finally, pulse 5 measures another half of the loop. Repeated domain switching is induced from pulse 1 to 5 in opposite directions before carrying out the measurements. Combining the data obtained from pulses 2 and 5, a complete $P$-$E$ loop comprising of contribution from both switchable and nonswitchable polarizations could be traced. Because of its structure, this set of pulses can be designated as Up Down Up Down (UDUD). Likewise, pulses 7 to 12 are used to measure the contribution from the nonswitchable component alone by repeatedly reinforcing the domains in the same direction. Combining the data obtained from the measurement pulses 8 and 11, another hysteresis loop could be constructed for nonswitchable polarization. The pulse set 7 to 12 can be termed as Up Up Down Down (UUDD). By subtracting the loop obtained by UUDD pulses from the one obtained by UDUD pulses, the intrinsic remanent hysteresis loop could be constructed (Fig. 4b). The entire train of twelve pulses is designed in such a way so that for measuring the contribution of both hysteretic and nonhysteretic polarization, two polarizing pulses (one square and another triangular) are employed prior to each measurement pulse consistently across the entire pulse train. For example, pulses 3 and 4 are used for switching the domains prior to measuring the contribution by pulse 5. Apart from sending much longer pulse train which induces repeated switching/reinforcement of the domains, remanent hysteresis measurement protocol yields the polarization loops differently from PUND. In PUND, positive halves of the loops comprising of both switchable and nonswitchable components are measured first while the negative halves are measured next. The remanent hysteresis protocol, on the other hand, measures the complete loop both in UDUD and UUDD sets. We demonstrate the efficacy of this protocol in extracting the intrinsic remanent polarization in an improper ferroelectric systems where the ferroelectric polarization is orders of magnitude smaller.

We used orthorhombic LuFeO$_3$ and Pr$_{0.55}$Ca$_{0.45}$MnO$_3$ samples vis-a-vis a standard displacive ferroelectric Pb(Zr$_{0.52}$Ti$_{0.48}$)O$_3$ for studying the utility of these techniques. In Fig. 1, we show the polarization versus field hysteresis loop obtained for LuFeO$_3$ at room temperature. Clearly, the loop is cigar-shaped commonly observed for lossy dielectric system where the charge $Q$ [= $\sigma.EAt$, $\sigma$ is the conductivity, $E$ is the elctric field, $A$ is the cross-sectional area and $t$ is the time] is directly related to the conductivity of the sample and not polarization. The polarization current-voltage characteristics obtained from this measurement yields a large leakage current background within which the domain switching polarization current is buried. The leakage current versus field plot is also shown in Fig. 1. In Fig. 2, we show the current and voltage profiles as well as the hysteresis loop obtained from PUND. The domain switching peak is visible in the polarization current versus field characteristics (Fig. 2c) yet the leakage current background is also quite strong. The hysteresis loop too does not resemble the one observed for a standard ferroelectric system such as Pb(Zr$_{0.52}$Ti$_{0.48}$)O$_3$. Fig. 2d shows the leakage current versus field plot. Finally, in Figs. 3 and 4, we show the remanent hysteresis loop as well as the corresponding voltage and current profiles. Remarkably, under identical conditions such as identical field, time scale, polarizing voltage etc., the loop for LuFeO$_3$ (Fig. 4b) turns out to be quite different and resembles the one observed in a standard ferroelectric sample. The domain switching peak in the polarization current-voltage characteristics is quite prominent in this case (Fig. 4c). The corresponding leakage current profile is shown in Fig. 4d.

We now discuss the underlying mechanisms of the PUND and remanent hysteresis protocols in order to examine their relative efficacy. The PUND protocol is being widely used in recent time.\cite{Feng, Kobayashi,Lunkenheimer-1,Lunkenheimer-2,Gich} As against the single triangular pulse used for measuring the polarization hysteresis loop for proper ferroelectric systems, this protocol helps in eliminating the nonhysteretic polarization. However, we show here that the pulse train used in PUND gives rise to limited switching of the domains and does not allow complete relaxation of both the hysteretic and nonhysteretic polarizations. In systems where intrinsic hysteretic polarization is small ($\sim$1-10 nC/cm$^2$), feroelectric domains coexist with nonferroelectric regions having variation in charge conduction characteristics. This inhomogeneous system with interfaces gives rise to depolarizing field as well and, therefore, relaxes over a much broader and distributed time scale. The incomplete relaxation of ferroelectric or nonferroelectric or even both the components yields a finite contribution which masks the actual relaxed ferroelectric polarization component. Therefore, a background current reminiscent of incomplete relaxation prevails. The switching kinetics of the polar domains too could follow a complicated pattern\cite{Noh} instead of well-known Kolmogorov-Avrami-Ishibashi model. All these are not adequately taken care of by the PUND protocol. The polarization and coercivity data obtained from PUND thus could still be inaccurate. On the contrary, the remanent polarization measurement protocol uses a train of fourteen pulses and repeated switching of direction of applied voltage. Across the entire train, each measurement pulse is consistently preceded by two polarizing pulses (one square and one triangular) which switch and reinforce the hysteretic and nonhysteretic components, respectively. The square pulse with infinitesimally small rise and decay time and the triangular pulse with steady rise and decay across a finite time scale ensure completion of switching/reinforcement of domains as switching and relaxation processes in such inhomogeneous systems could take place across a broader time span with wide variation in characteristic time scales. The rationale behind employing this train of pulses stems from the facts such as (i) normal distribution of switching voltage of the domains,\cite{Evans} (ii) complicated switching/relaxation kinetics of the domains,\cite{Kim,Noh} (iii) repeated switching\cite{Johann} ensuring measurement of contribuion from stable domains alone. The entire time scale of the measurement and this additional switching/reinforcement of domains ensure rather complete relaxation of both the switchable and nonswitchable components of the polarization. Of course, as mentioned later, a guiding principle for choosing the time scale of the voltage pulses exists for this protocol as well. This process, nevertheless, eventually yields the fully relaxed intrinsic ferroelectric polarization. Comparison of the data obtained from PUND and remanent hysteresis protocol at different time scales for low resistive LuFeO$_3$ sample clearly establishes this point (Fig. 5). While the polarization loop obtained from PUND shows time dependence, the one obtained from the remanent hysteresis protocol is time independent. It is important to point out here that even in proper ferroelectric BaTiO$_3$, one observes significant relaxation of polarization across a longer time scale in thin film samples because of large depolarizing field $E_d$ generated at the sample-electrode interface.\cite{Kim} For this case too, an appropriate measurement protocol needs to be employed to ensure accurate determination of remanent polarization and coercivity.   

\begin{figure}[ht!]
\centering
\includegraphics[width=70mm]{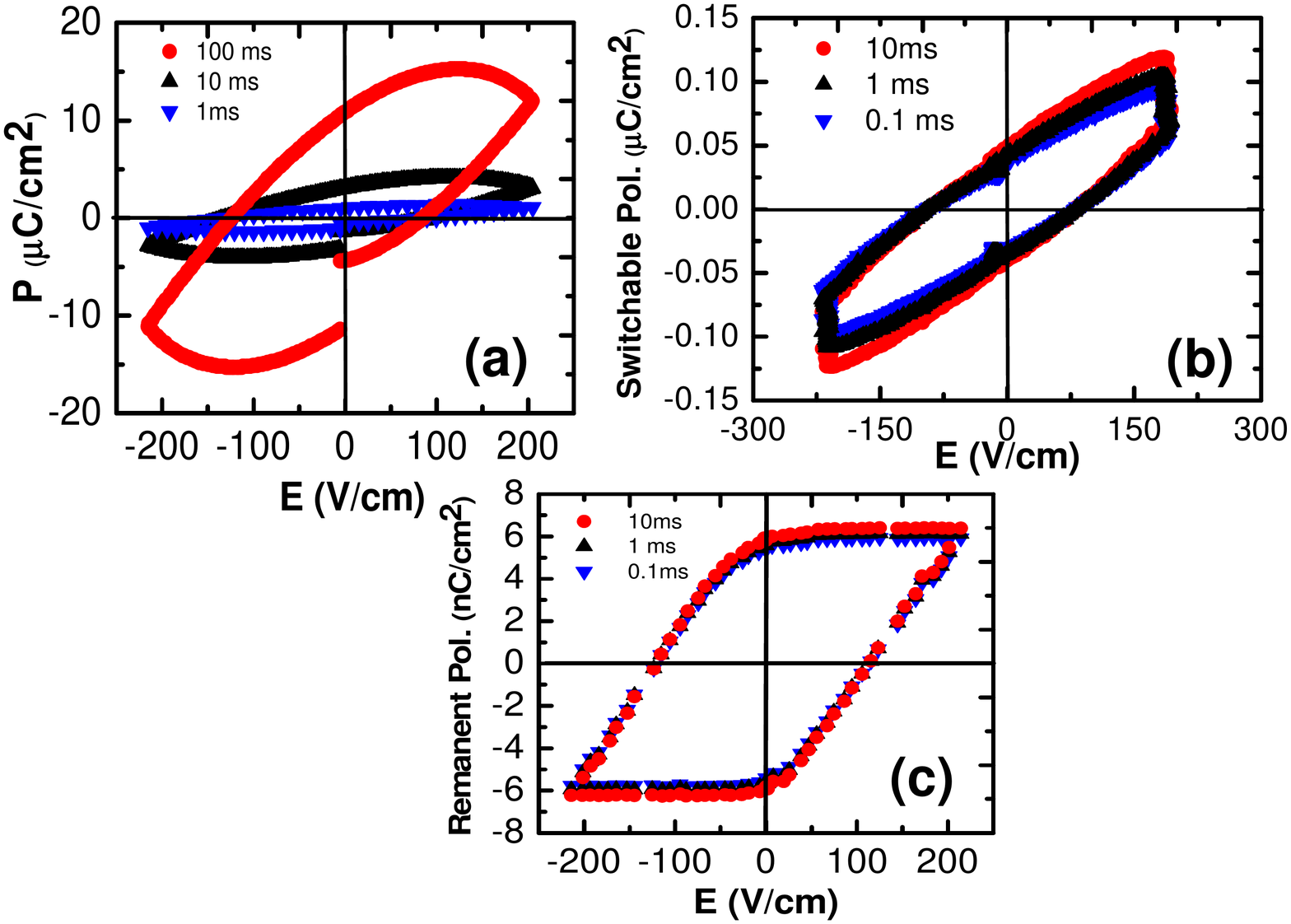}
\caption{(color online) The time scale dependence of polarization versus electric field hysteresis loops obtained from (a) single triangular pulse, (b) PUND, and (c) remanent hysteresis protocols for orthorhombic LuFeO$_3$. }
\label{overflow}
\end{figure}

\begin{figure}[hb!]
\centering
\includegraphics[width=80mm]{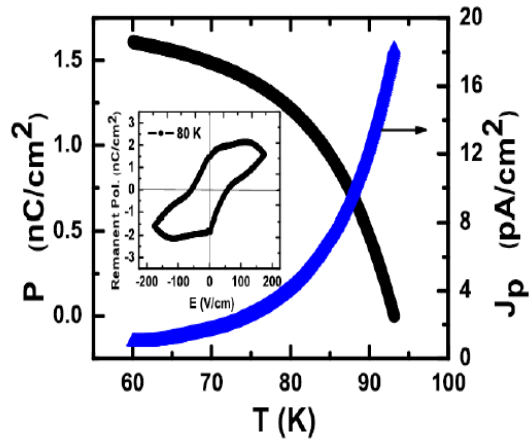}
\caption{(color online) The pyroelectric current and polarization obtained for orthorhombic Pr$_{0.55}$Ca$_{0.45}$MnO$_3$ as a function of temperature; inset shows the remanent hysteresis loop. }
\label{overflow}
\end{figure}

We also compare these results with what is obtained from pyroelectric current technique. In this technique, the remanent polarization is measured by bringing down the sample from above transition temperature under a dc bias. The bias field is then switched off and the current leads are shorted to eliminate space charge. Finally, the temperature is raised at a fixed rate (1 K/min) without any bias and the pyroelectric current $I_p(T)$ is measured as a function of temperature. Integration of $I_p(T)$ yields the polarization $P(T)$. However, it has already been pointed out\cite{Feng} that pyroelectric current technique yields errorneous polarization because of inadequate poling in low-resistive samples. The poling field is applied at above the transition point. Because of lower resistivity at that temperature, the poling field applied cannot ensure complete saturation of polarization at a lower temperature at which the actual measurement takes place. The space charge also influences the polarization obtained from this technique in low-resistive samples.\cite{Feng} We have used pyroelectric current technique to measure the polarization in orthorhombic Pr$_{0.55}$Ca$_{0.45}$MnO$_3$ (Fig. 6). The magnitude of the polarization obtained from pyroelectric current is lower than that obtained from remanent hysteresis (inset of Fig. 6) indicating incomplete saturation of polarization. Therefore, prior determination of the field required to achieve complete saturation for such low-resistive samples is essential.

Based on the above measurements, it is possible to establish the utility of a particular technique for measuring the intrinsic ferroelectric polarization of different samples depending on their leakage and ferrolectric polarization.\cite{supplementary} For proper ferroelectric systems such as BaTiO$_3$ or Pb(Zr$_{0.52}$Ti$_{0.48}$)O$_3$ with very high resistivity ($>$10$^{9}$ $\Omega$cm) and polarization ($>$10 $\mu$C/cm$^2$), conventional single triangular pulse is appropriate as the contribution from non-hysteretic and leakage components is quite negiligible. Comparison of the results obtained for such samples from different techniques establishes this point. For organic or improper ferroelectric systems with low resistivity ($\sim$10$^7$-10$^9$ $\Omega$cm) and polarization ($\sim$0.1-1 $\mu$C/cm$^2$), PUND technique does offer intrinsic switchable polarization while simple triangular pulse yields a loop reminiscent of lossy dielectric systems provided the PUND protocol ensures complete relaxation of both the hysteretic and nonhysteretic components. In fact, for those samples, comparison of the results obtained from PUND and remanent hysteresis protocol shows nearly comparable ferroelectric polarization. In the case of the samples of even lower resistivity ($\le$10$^6$-10$^7$ $\Omega$cm) and polarization ($\sim$1-10 nC/cm$^2$), however, it appears that only the remanent hysteresis measurement protocol could measure the intrinsic ferroelectric polarization. In this context it needs to be mentioned that, like other protocols, following two important issues should be taken care of in order to ensure appropriate measurement of intrinsic hysteretic polarization - (i) quality of sample-electrode interface and (ii) time scale of the measurement. The sample-electrode interface should be sharp for perfect electrostatic screening of the field right at the interface. Absence of finite screening length ensures zero depolarizing field and hence accurate measurement of the polarization of the sample. The time scale of the voltage pulse, on the other hand, should be comparable to the time constant $\tau$ of the circuit ($\tau$ = $R.C$; $R$ = resistance, $C$ = capacitance) and the time scale of intrinsic polarization switching kinetics.\cite{Li} In systems where $\tau$ governs the switching process strongly, the time scale of the applied voltage pulse should be smaller than $\tau$ for ensuring appropriate measurement of polarization switching and the magnitude of polarization. For samples with still lower resistivity and polarization, a similar yet longer pulse train could be designed which induces repeated switching of the polarization and ensures accurate determination of intrinsic remanent polarization. In fact, further work needs to be done on correlation among the time scale of measurement, shape of the voltage pulse, number of pulses, polarization switching kinetics, and accurate estimation of the intrinsic polarization for samples with even smaller resistivity and intrinsic polarization.          
 
This work is supported by DST, Govt of India. One of the authors (UC) acknowledges support in the form of Research Associateship from CSIR. Another author (SG) acknowledges support in the form of Senior Research Associateship from CSIR.

\end{document}